\documentstyle[preprint,aps]{revtex}
\begin{document}
\draft

\title{Anomalous Proximity Effect in Underdoped YBa$_2$Cu$_3$O$_{6+x}$ Josephson
Junctions}
\author{R. S. Decca\cite{address},$^{1,2}$, H. D. Drew$^1$, E. Osquiguil$^3$, B. Maiorov$^3$, and J. Guimpel$^3$}
\address{$^1$Laboratory for Physical Sciences, 
and Department of Physics, University of Maryland, College Park, MD 20742,\\ 
$^2$ Department of Physics, Indiana University Purdue University Indianapolis, 402 N. Blackford St., 
Indianapolis, IN 46202,\\
$^3$Centro At\'omico Bariloche and Instituto Balseiro, 8400 S. C. de Bariloche, R. N., Argentina.}
\date{\today}
\maketitle

\begin{abstract}
Josephson junctions were photogenerated 
in underdoped thin films of the YBa$_2$Cu$_3$O$_{6+x}$ family using a near-field scanning 
optical microscope. The observation of the Josephson 
effect for separations as large as 100 nm between two wires indicates the existence of an 
anomalously large proximity effect and show that the underdoped insulating material in the gap 
of the junction is readily perturbed into the superconducting state. The critical current of 
the junctions was found to be consistent with the conventional Josephson relationship. This 
result constrains the applicability of SO(5) theory to explain the phase diagram of high 
critical temperature superconductors.
\end{abstract}
\pacs{74.50.+r, 74.80.Fp, 74.72.Bk, 74.76.Bz}

%\narrowtext
%\twocolumn

Despite being among the most intensely studied condensed-matter systems, high temperature 
superconductors (HTS) have resisted a microscopic understanding\cite{phase}. They are 
strongly anisotropic highly correlated electronic systems, showing anomalous characteristics in
both the superconducting and non-superconducting phases. In particular, the
nature of the transition between the low carrier concentration insulating
antiferromagnetic (AF) phase and the high carrier concentration metallic and
superconducting (SC) phase is not known and it is believed to be key
to uncovering the superconducting mechanism in these materials. The quest to gain
a better understanding of these issues is reflected in extensive
experimental and theoretical work\cite{phase,so5,fish,oren,gap,demler,jj5,last}. Recently 
an elegant theory, based on SO(5) group symmetry\cite{so5}, proposed a basic
framework to explain the HTS phase diagram. A five component superspin was
introduced with two of its components associated with the order parameter in a {\it d}-wave
SC state and the other three identified with the order parameter of the
AF phase. In this theory, the quantum phase transition between the AF and SC
phases corresponds to a change in the orientation of the superspin in this five-dimensional space. 
A different
approach for describing the rich phase diagram of the HTS materials postulates 
superconducting pairing at a temperature $T^*$ well above the
superconducting critical temperature $T_c$. The low stiffness of the
superconducting order parameter, due to their large penetration length\cite{fish}, leads to 
fluctuations in the phase of the order parameter\cite{oren,gap} between $T^*$ and 
$T_c$. It is not until phase coherence is
achieved at $T_c$ that superconductivity is established in the material. The
overall behavior of the superconductor between $T^*$ and $T_c$ resembles
that of a Kosterlitz-Thouless transition in conventional
two-dimensional superconductors.

The electronic nature of the underdoped system near the superconducting 
state is significantly different than the normal state found in
conventional superconductors. Consequently, the experimental manifestation of 
superconductivity may also be expected to differ. Within
the framework of the SO(5) theory, Demler {\it et al.}\cite{demler} have predicted 
that the current-phase relationship in the
coupling between two HTS separated by a thin AF layer (a {\it SAS}
junction) is modified from the Josephson relation $I_J = I_o sin\phi$, with $\phi$ the 
superconducting phase difference across the junction\cite{Tinkham,degennes}. 
If the thickness of the junction $d < d_c = \pi \xi_A$ the AF material becomes a
superconductor for $\phi < \phi_c$ and ``conventional'' Josephson effect occurs only for 
$\phi > \phi_c$, where $\xi_A$ represents a new superconducting correlation 
length in the AF phase\cite{demler} and 
$\phi_c \sim \pi \sqrt{1-(\frac{d}{d_c})^2}$\cite{jj5}.

This difference in the current-phase relationship of a {\it SAS} junction
leads to different behavior of Josephson junctions, as investigated
by den Hertog and co-workers\cite{jj5}. In particular, they predicted that the critical current 
of the junctions $I_c(H)$ would show a linear decrease with an applied magnetic field around 
$H=0$\cite{jj5}; 
i. e., a cusp in the $I_c(H)$ dependence, instead of the quadratic low field behavior found in 
conventional junctions. We are not aware of an analogous prediction within the framework of the
fluctuating-phase model. 

Both the fluctuating-phase approach and the SO(5)
theory share the possibility of a large proximity effect when the
material separating the HTS wires is the insulating HTS precursor. These considerations suggest 
that Josephson junctions may be made by separating HTS with a
relatively thick layer of the precursor material\cite{fish,demler}.

We exploit the capability of locally photodoping an
insulating {\it R}Ba$_2$Cu$_3$O$_{6+x}$ material (with {\it R} a rare
earth), to induce superconducting wires separated by a non-superconducting
region\cite{us3}. The flexibility provided by our near-field scanning optical microscope 
(NSOM) allows us to vary the gap between the $w \sim 150$ nm wide superconducting wires.

The samples under consideration are {\it c}-axis oriented thin films. One is
a 180 nm thick GdBa$_2$Cu$_3$O$_{6+x}$ film grown on (100) MgO substrates by
dc-magnetron sputtering. The other sample is a 120 nm thick YBa$_2$Cu$_3$O$_{6+x}$ (YBCO) film 
deposited on a SrTiO$_3$ substrate by laser ablation.
The as-grown films show good physical properties with linear temperature
dependence of the dc resistivity. Their critical temperatures, determined by
ac-susceptibility, were $T_c = 89.4$ K and $89.2$ K, respectively. To place the samples in the 
insulating side of the Metal-Insulator transition (MI), their
oxygen content was adjusted to $x \sim 0.4$\cite{osqui}. After reduction, the resistivity of 
the samples at 4 K was found to be $\rho \sim 6$ m$\Omega$cm. The results obtained in both 
samples are very similar and we will concentrate on the data obtained in the YBCO film.

The film was mounted on the insert of a continuous-flow He cryostat and
photodoped with an Al-coated 60 nm aperture NSOM probe\cite{us3}. Light from either a 3 mW 
He-Ne or a 1 mW, $\lambda$ = 1.55 $\mu$m, InGaAsP laser was coupled into the inputs of a 50/50
2 $\times$ 2 optical fiber coupler. One of the outputs of the coupler was connected to the
NSOM probe, while the other was used to monitor the laser stability.
Photogeneration was accomplished by illuminating with the 1.96 eV light from the He-Ne laser, 
which is close to the maximum of the photodoping efficiency\cite{kudinov}. The photoinduced 
changes in the sample were detected by imaging the 
reflectance variations at $\lambda$ = 1.55 $\mu$m with the InGaAsP laser.
The reflected light was collected in the far field with conventional optics\cite{us2}. The 
InGaAsP laser was chosen because $\lambda$ = 1.55 $\mu$m radiation provides the maximum 
change in reflectivity when crossing the MI transition\cite{orenstein} and it does not
induce any further photogeneration\cite{us3,kudinov}. Photon fluxes per unit
time were estimated to be $Q_{exc} \sim 8.6 \times 10 ^{20}$ {photons/(cm$^2$
s) and $Q_{ref} \sim 3.2 \times 10 ^{20}$ photons/(cm$^2$s) for
photoexcitation and reflectivity measurements, respectively. All the
measurements involving the NSOM were performed at room temperature. When
necessary to prevent the superconducting wires from decaying by {\it e-h}
recombination\cite{us3,kudinov,osqui2}, the NSOM head was removed, the
cryostat closed and pumped to 10$^{-6}$ Torr and the sample cooled to 200 K
in less than 15 min\cite{recombination}. It is well documented\cite{us3,osqui2} 
that below 250 K the photoinduced state is metastable. 

Typical NSOM reflectance scans are shown in Fig. \ref{wire}. The wires were
defined and the scans obtained as described in Ref. \cite{us3}. Fig. \ref
{wire} shows that the reflectance and $T_c$ of the wires increase with the duration of the 
photogeneration.
These results are explained by the photoinduced local increase of free
holes in the CuO planes of YBCO\cite{us3,kudinov,osqui2}.

Josephson junctions were defined by photogenerating a wire and leaving an
unilluminated gap along its length, as described in Ref. \cite{us3}. A
typical example of these junctions is illustrated in Fig. \ref{junction}a.
We determine the gap $d$ between the superconducting wires from the reflectance
data shown in Fig. \ref{wire}. We define $d$ as the range where the
reflectance is lower than that corresponding to the wire in Fig. \ref{wire}b, which has 
a $T_c \sim
$ 4 K. From this definition of $d$ the part of the junction between points {\bf b} of Fig. 
\ref{junction}b is insulating in character. The separation between these points is $\sim 90$ nm. 
Because of the finite resolution of the NSOM this procedure gives a
lower limit of the thickness of the barrier between the wires. After
defining an identical junction and cooling the system to 4 K, $I-V$
characteristics were obtained, as shown in the inset of Fig. \ref{junction}a. The zero 
dissipation region shows the existence of Josephson effect
between the two wires\cite{Tinkham}. The rounding of the I-V curves is
understood in terms of thermal fluctuations in the Josephson junction\cite{Tinkham}.

The observation of the Josephson effect for a separation between
superconducting wires much
greater than the coherence length in the superconducting state
($\xi_o \sim 1$ nm\cite{phase}) is one of the main results of this
paper. This ``colossal'' 
proximity effect is inexplicable even if it is assumed that the material
in the gap is
metallic. In this case, the coherence length for the metal in the clean
limit is $\zeta \simeq
\frac{\hbar v_f}{2 \pi k_B T}$, where $v_f$ is its Fermi velocity{\it
\cite{Tinkham,degennes}}.
For reasonable values of $v_F$, $\zeta$
turns out to be the same order of magnitude as $\xi_o$. From
these considerations we conclude that conventional proximity effect
cannot explain our results and that the insulating material in the junctions
exhibits an anomalously large proximity effect. Further evidence that a
superconducting state is induced in the gap material is provided by the
value of the critical
current of the junction. For the  $d = 45$ nm junction, $I_c = 2.6
\mu$A, very close to the
measured value of $I_c = 11.5 \mu$A for the wire of Fig. \ref{wire}a.

As can be seen in the inset of Fig. \ref{junction}b,
$I_{c}R_{N}$ ($R_{N} = \frac{dV}{dI}|_{I = 7\mu{\rm A}}$ is the shunt
resistance of the junction) is nearly independent of $d$ for $d < 110$
nm. This
fact has implications for both SO(5) theory and the fluctuating-phase
models: When the existence of free vortices is used to explain the lack
of
phase coherence in the gap material, $I_{c}R_{N}\propto \exp (-d/2\zeta
_{g})$ is expected,
where $\zeta _{g}\simeq \xi _{o}\exp (T_{\Theta }/T)$ is the
correlation length for phase fluctuations, and $T_{\Theta }$ is the
phase
stiffness expressed in temperature units\cite{oren}. From the
observed weak $d$ dependence of
$I_{c}R_{N}$ when $d \stackrel {\textstyle <}{\sim}$ 100 nm, it follows
that
$\zeta _{g}\gg 90$ nm and $T_{\Theta }\stackrel{\textstyle >}{\sim} 30$ 
K. As the separation 
between wires increases, however, a collapse in the phase coherence is 
observed and no 
Josephson effect is observed. In the SO(5) theory $I_{c} \propto \exp 
(-d/\xi _{A})$ implying that the correlation length satisfies 
$\xi _{A}\gg d$\cite{demler,Tinkham}. It is possible to have
strong $d$ dependence in both $I_c$ 
and $R_N$, that cancel each other, as in the case in a tunnel Josephson
junction in 
conventional superconductors. This possibility is ruled out, however, by
the measured temperature 
dependence of the resistance of the non-superconducting material. $R_N$
is temperature 
independent {\it if the Josephson effect is 
observed}. Once the separation between the wires is large enough that
the Josephson effect is absent, an insulating-like response is observed for
the unphotodoped material. The difference found in the temperature
dependence of
the non-superconducting material for different junctions is attributable
to their difference in length, as shown in Fig. \ref{resistance}. The minor
deviations between the data sets might be associated with the specific
geometric details of the path followed by the transport current in each
case, together with the definition of $d$.

The observed behavior of these Josephson junctions may be
expected if the gap material {\it is thermodynamically very close to the
superconducting state} and the presence of the superconducting leads
quenches
the superconducting phase fluctuations in the gap material. The
induction of superconductivity in
the non-superconducting material is a feature of both the SO(5)
theory\cite{so5,demler} and the fluctuating phase models{\it
\cite{fish}}.
More generally, as discussed in
Ref. \cite{demler}, the only ingredient necessary to obtain such a
large correlation length is the
close proximity to a second order quantum phase transition.

One observation of this experiment appears to be in contradiction with the
SO(5) theory. Since the observed behavior of $I_cR_N$ implies that $\xi_A \gg d$ our experiments 
correspond to the case analyzed in Ref. \cite{demler}, {\it if the material between the 
superconducting wires is in the AF phase}. In this scenario, $I_c(H)$ should show a cusp 
for $H \rightarrow 0$\cite{jj5}. The cusp is expected even for situations where the critical current is 
not uniform along the width of the junction, as expected for junctions made by our
technique\cite{expla}. The results obtained for small magnetic fields are shown in Fig. 
\ref{smallh}. The figure shows that for $H \rightarrow 0$, $I_c(H) \not\propto |H|$, but is 
better described by a quadratic dependence. Also shown in the figure is the calculated 
$I_c(H)$ using the model from Ref. \cite{jj5}.  The experimental value of 
$H = 135$ Oe for the first flux-quantum trapped in the 
junction, obtained from the Fraunhoffer-like dependence of $I_c(H)$, was used in the calculation. 
As shown in the figure, {\it the agreement between the experimental data and the calculation is reasonable only if} 
$d \sim d_c$, in clear contradiction
with the $\xi_A \gg d$ conclusion obtained from Fig. \ref{junction}b. 

The effects observed in the junctions are independent of the sequence in
which the magnetic field is applied. The same results are observed if the
sample is cooled through the superconducting transition in zero magnetic
field, and then the field is brought up to the desired value (zero-field
cooled experiment); or if the sample is cooled in the desired field
(field-cooled experiment). Although the lower critical field in these underdoped HTS materials 
is expected to be very small, no difference is found since 
$w < \lambda_p \sim 500$ nm\cite{lambda}, where $\lambda_p$ is the penetration
length in the superconductor. In this condition size effects are very important and it is 
energetically unfavorable to
induce vortices in the superconducting wires.

The results described in this report show that the local combination of superconducting and insulating 
materials, obtained by photodoping
insulating YBCO with a NSOM probe, provides an ideal opportunity to examine the validity of the 
different models of high critical temperature superconductors. The observation of the Josephson 
effect for separations $d \sim$ 100 nm  cannot be 
explained by the conventional
proximity effect. We conclude that the material between the superconducting
wires, although insulating, must be very close to a superconducting phase
transition since superconductivity with a large phase coherence length can be induced in it. We
have also been able to restrict the margin of applicability of the SO(5)
theory to explain the phase diagram of HTS. In view of the observed results
and with the available models, only the existence of an AF and SC phases
separated by a quantum disordered phase remains compatible with the theory.
Although our results seem to be in qualitative agreement with
fluctuating-phase models, more theoretical work is required to obtain the
characteristics of the Josephson effect within this picture. 

We would like to thank C. Lobb, M. P. A. Fisher, and D. Prober for useful discussions. We are 
also indebted
to A. J. Millis, S. A. Kivelson, and E. Demler for a critical reading of the manuscript. This 
work was partially
supported by the National Science Foundation's MRSEC program through the
University of Maryland/NSF grant DMR-963252. Work at the Centro At\'omico Bariloche and 
Instituto Balseiro was suported by grant ANPCYT PICT97-03-00061-01117 and grants by Consejo 
Nacional
de Investigaciones Cient\'{\i}ficas y Tecnol\'ogicas (CONICET), Fundaci\'on
Antorchas, and Fundaci\'on Balseiro of Argentina. J. G. and E. O. are
members of CONICET, Argentina.

\begin{figure}[tbp]
\caption{(color) Reflectance of photodoped wires. The wires were photodopped for 
{\bf (a)} 5000 s, {\bf (b)} 1300 s, {\bf (c)} 800 s, and {\bf (d)} 600 s.
The number in each 1.24 $\times$ 0.64 $\mu$m$^2$ image indicates the
percentile increase in reflectance of the wire (at $\lambda = 1.55 \mu$m) with respect 
to the unphotodoped material. The coordinate system used throughout the
paper is shown in {\bf (a)}. The inset shows the temperature dependence of the
resistance for each wire, normalized by the value of the resistance at 15 K. }
\label{wire}
\end{figure}

\begin{figure}
\caption{(color) Josephson effect in photogenerated junctions. {\bf (a)} 1.24 $\times$ 0.64 $\mu$m$^2$ 
reflectance data of a photogenerated junction, normalized to the reflectance of the 
unphotodoped material. The gap in the illuminated region was $D$ = 220 nm. The inset
shows the $I-V$ characteristic curve. {\bf (b)} Line cut along the dotted
line in {\bf (a)}. The letters represent the different intensities observed in
Fig. \ref{wire}. The inset shows the value of $I_c R_N$ for junctions fabricated with different 
$d$.}
\label{junction}
\end{figure}

\begin{figure}
\caption{Temperature dependence of the resistance of the junctions that do not show Josephson 
effect. The inset
shows the data normalized by the separation between the superconducting wires.}
\label{resistance}
\end{figure}

\begin{figure}
\caption{Magnetic field dependence of the critical current close to $H$ = 0
Oe. For comparison, $I_c(H)$ obtained from Ref. [7] for different values of 
$\delta = \frac{d}{d_c}$ is included. In all cases the curves have been normalized for $H = 0$ 
Oe.}
\label{smallh}
\end{figure}


\begin{references}
\bibitem{recombination}  The recombination time of different samples at room temperature has been 
found to vary between 30 min and more than a day. Although the large dispersion in 
recombination time is not understood, it is known that sample
morphology, growth methods, and sample thickness, among other factors,
affect it. In our case, the recombination time for the YBCO sample was $\sim 
$ 10 hrs, while the GdBa$_{2}$Cu$_{3}$O$_{6.4}$ sample had a recombination
time $\sim $ 50 min.
\bibitem{expla}  In this case, $I_{c}(H)=\frac{1}{L}\int_{0}^{L}I_{c}(\phi
(x))dx$, where $\phi $ is position dependent through the gauge-invariant
superconducting phase difference across the junction and the spatial
dependence induced by the particular details of the geometry of the
junction. A numerical integration yields $I_{c}(H) \propto |H|$. This result
is to be expected, since it is obtained by adding a heavily weighted
``anomalous'' junction with a lightly weighted conventional one.
\end{references}
\end{document}